\documentclass[aps,pre,bibtex,twocolumn,notitlepage,superscriptaddress]{revtex4-2}

\usepackage{epsf,graphicx,amssymb,amsmath,color}
\usepackage[latin1]{inputenc}

\usepackage[squaren, Gray, cdot]{SIunits}
\usepackage{color}
\usepackage[francais]{babel}
\usepackage{hyperref}

\hypersetup{colorlinks=true,linkcolor=blue,citecolor=blue}

\newcommand{\ve}[1][K]{\mathbf{#1}}

\begin{document}

\title{Decay  dynamics of a single spherical domain in near-critical phase-separated conditions}

\author{Raphael Saiseau}
\email{r.a.saiseau@utwente.nl}
\affiliation{Univ. Bordeaux, CNRS, LOMA, UMR 5798, F-33400, Talence, France.}
\affiliation{Physics of Fluids, University of Twente, 7522 NB Enschede, The Netherlands.}

\author{Henri Truong}
\affiliation{Univ. Bordeaux, CNRS, LOMA, UMR 5798, F-33400, Talence, France.}
\affiliation{Univ. Bordeaux, CNRS, CRPP, UMR 5031, F-33600, Pessac, France.}

\author{Thomas Gu\'erin}
\affiliation{Univ. Bordeaux, CNRS, LOMA, UMR 5798, F-33400, Talence, France.}

\author{Ulysse Delabre}
\affiliation{Univ. Bordeaux, CNRS, LOMA, UMR 5798, F-33400, Talence, France.}\

\author{Jean-Pierre Delville}
\email{jean-pierre.delville@u-bordeaux.fr}
\affiliation{Univ. Bordeaux, CNRS, LOMA, UMR 5798, F-33400, Talence, France.}

\begin{abstract} 
Domain decay is at the heart of the so-called evaporation-condensation Ostwald-ripening regime of phase ordering kinetics, where the growth
of large domains occurs at the expense of smaller ones, which are expected to `evaporate'.
We experimentally investigate such decay dynamics at the level of a single spherical domain picked from one phase in coexistence and brought into the other phase by an opto-mechanical approach, in a near-critical phase-separated binary liquid mixture. We observe that the decay dynamics is generally not compatible with the theoretically expected surface-tension decay laws for conserved order parameters. 
Using a   mean-field description, we  quantitatively explain this apparent disagreement by the gradient of solute concentrations induced by gravity close to a critical point. Finally, we  determine the conditions for which buoyancy becomes negligible compared to capillarity and perform  dedicated experiments that retrieve the predicted surface-tension induced  decay exponent. 
The surface-tension driven decay dynamics of conserved order parameter systems in the presence and the absence of gravity, is thus established at the level of a single domain. 
\end{abstract}

\maketitle

Phase ordering kinetics is an ubiquitous and fundamental process \cite{onuki2002phase,Bray1994}, that  characterizes the irreversible evolution of a system initially out-of-equilibrium into well separated coexisting phases. 
For conserved order parameter systems, the nucleated domains usually evolve from an intermediate free-growth regime to the so-called late stage Ostwald ripening \cite{marqusee1983kinetics,taylor1998ostwald,ratke2002growth}, where the larger domains grow at the expense of the smaller ones which decay, or ``evaporate'' due to interfacial tension, as nicely demonstrated by the classical evaporation-condensation LSW theory of Lifshitz, Slyozov \cite{lifshitz1961kinetics} and Wagner \cite{wagner1961ostwald}. 
This very general phenomenon is at work in all phase transition dynamics involving conservation of the order parameter and has consequently huge implications in out-of-equilibrium liquids, soft matter and material sciences \cite{alloyeau2010ostwald,jiao2003ostwald,Ardell2005}. 

The decay of the smallest domains was clearly observed in two dimensional systems, such as metallic islands \cite{morgenstern2001decay} or during a liquid-solid transition triggered in confined geometry \cite{krichevsky1995ostwald} in order to illustrate the importance of correlations between evolving domains. In three dimensions, its investigation is essentially indirect as most coarsening experiments, performed at the scale of a large number of interacting droplets, and measured by scattering techniques \cite{Cumming1992} or by direct visualization of droplet assemblies \cite{snyder2001transient,alkemper1999dynamics}, focus instead on the evolution of the droplet distribution and its comparison to LSW predictions. 
So, although at the basis of the evaporation-condensation Oswald-ripening, the  surface-tension driven evaporation of small droplets in conserved  order parameter systems remains almost unexplored. Conversely, standard experiments \cite{fuchs1959evaporation,Frohn2000Book}
 on the evaporation of single liquid drops report the classical $R$-squared law, in which the drop radius $R$ evolves with time $t$ as $R\sim(t_{\mathrm{f}}-t)^\alpha$ with $\alpha=1/2$, where $t_{\mathrm{f}}$ is the final evaporation time. The involved mechanism is nonetheless not driven by surface tension for  scalar  conserved systems, for which theory \cite{Bray1994} predicts instead $\alpha=1/3$ (unlike non-conserved systems, for which $\alpha=1/2$ for surface-tension driven decay \cite{Bray1994,krapivsky2010kinetic}).
 This exponent $\alpha=1/3$ has been observed in the different context of the thinning dynamics of liquid necks \cite{lo2019diffusion,aagesen2010universality}.
However, surprisingly,
this domain decay prediction at the scale of a single spherical drop has never been confronted to experiments, whereas it plays a key role in evaporation-condensation processes. 
This is the subject of the present letter.

Starting from a near-critical phase-separated liquid mixture at equilibrium to keep the universality of Ising models, we use the optical radiation pressure of a laser wave to destabilize the meniscus separating the two phases in equilibrium
and produce a size-controlled single drop of one phase immersed into the other, at a chosen altitude. Then, we turn off the laser and look at the drop evaporation. 
While we expected the $R\sim (t_{\mathrm{f}}-t)^{1/3}$ scaling, 
our observations firmly indicate a richer dynamics due to  gravity \cite{ratke1985influence, kumaran1998effect}. We build a dedicated mean-field model which  incorporates gravity and confront our measurements to predictions. Finally, we `close the loop' 
by determining from the modeling the conditions for which buoyancy becomes negligible compared to capillarity and performing dedicated experiments that retrieve the expected gravity-free decay exponent $\alpha=1/3$. We thus characterize the surface-tension driven decay dynamics of conserved systems, at the level of a single spherical domain.   

\begin{figure}[ht!]
    \includegraphics[width=\columnwidth]{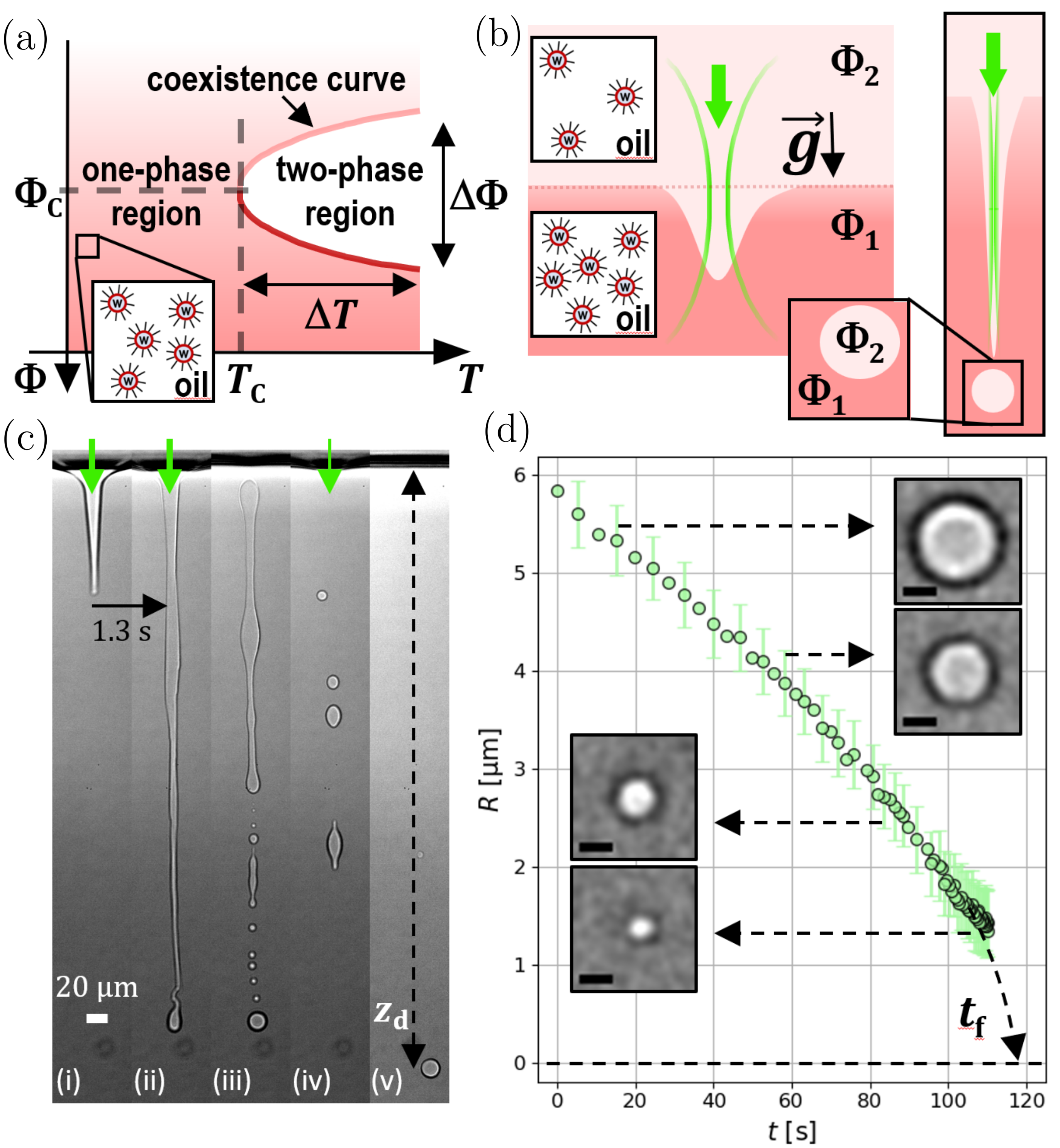}
    \caption{ (a) Schematic phase diagram of the critical micellar solution, with $\Phi$ the micellar volume fraction, and ($T_{\textrm{c}},\Phi_{\textrm{c}})$ the coordinates of the critical point. A cartoon of a single water-in-oil micelle, of typical size 4~nm, is also shown (inset). 
    (b) Schematics of the optical bending and jetting instability of the meniscus ($T>T_{\textrm{c}}$) induced by a laser beam focused at the meniscus (green). 
    (c) Typical image sequence of the optical jetting instability. We show (i, ii) the jet formation, (iii) the Rayleigh-Plateau instability  when the laser is turned off the first time, (iv) the radiation pressure effect  at lower laser power to force droplet coalescence, and (v) the resulting final evaporating droplet optically pushed to a given altitude. (d) Decay dynamics for $\Delta T= 4$~K, including snapshots, of the produced droplet when the laser is permanently 
     turned off. $t_\mathrm{f}$ is the first time when the droplet cannot be detected any longer. Scale bars: $5~ \mathrm{\mu m}$. }
\label{fig:1}
\end{figure}

\textit{Experimental system.} 
We consider a water-in-oil micellar solution of critical composition (water, $9 \%$~wt, toluene, $79 \%$~wt, SDS, $4 \%$~wt, butanol, $17\%$~wt) constituted of water nanodroplets coated by surfactant (the micelles, of radius $r\simeq 4$nm) homogeneously dispersed in an oil continuum (the toluene) \cite{saiseau2022near,Petit2012}, see the Supplemental Material  for details~\cite{SM_CriticalShrinking}. As a pseudo binary liquid mixture, where the micelles behave as the solute, it shows an Ising ($d=3, n=1$) low critical point at $T_{\mathrm{C}}\simeq 38^{\circ}~$C, see Figure \ref{fig:1}({a}).  
The microemulsion is contained inside a tight fused-quartz cell of 2 mm thickness thermally-controlled with a PID in a homemade brass oven. 
When $\Delta T\equiv T-T_{\mathrm{C}}>0$, the solution separates into two phases of respective micellar volume fractions $\Phi_{1}$ and $\Phi_{2}$ (with $\Phi_{2}<\Phi_{1}$), separated  at equilibrium by an horizontal  interface, called hereafter the meniscus. To produce thermodynamically metastable droplets, we use the optical radiation pressure of a continuous laser wave focused at the meniscus to induce a liquid jet \cite{girot2019conical,casner2003laser} of the  phase $\Phi_2$ into the  phase $\Phi_1$, see Figure \ref{fig:1}({b}). When the laser is turned off, drops are produced through the Rayleigh-Plateau instability \cite{Petit2012}. Then, using again the radiation pressure at lower beam intensity, we force some of these drops to  coalesce into a single one which is further pushed opto-mechanically to a chosen altitude, see Figure \ref{fig:1}({c}). Thermodynamically out-of-equilibrium due to its finite size, the produced droplet immediately starts to evaporate while buoyancy makes it rise toward the meniscus.

The droplet decay is captured using $\times 20$ or $\times 50$ long working distance Olympus\textsuperscript{\textregistered} microscope objectives and a Phantom\textsuperscript{\textregistered} VEO340L camera. 
A custom-made ImageJ\textsuperscript{\textregistered} code is used to detect the intensity gradient maxima at the drop edges to extract the droplet radius $R$, and we checked the calibration with other detection methods, see SM~\cite{SM_CriticalShrinking}.
The altitude  $z_{\mathrm{d}}$  of the center of mass of the droplet is also recorded, with the convention that $z_{\mathrm{d}}$ increases in the downward direction, and $z_{\mathrm{d}}=0$ at the meniscus. An example of droplet evaporation dynamics is shown on Figure \ref{fig:1}({d}).
 
 \begin{figure}[ht!]
    \includegraphics[width=\columnwidth]{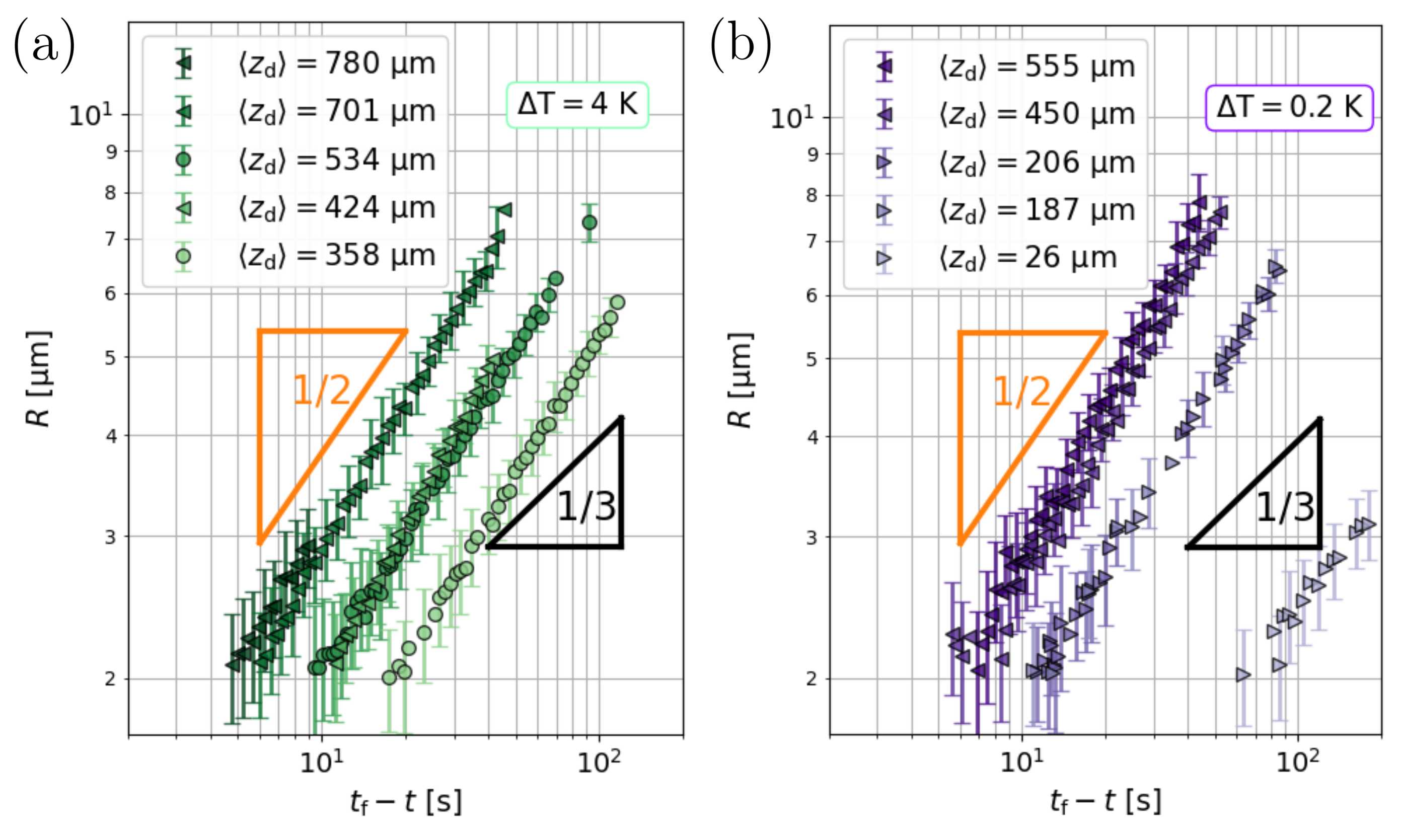}
    \caption{Droplet radius $R$ as a function of $t_{\mathrm{f}} - t$  for (a) for $\Delta T = 4~$K and (b) $\Delta T = 0.2~$K.
The legend indicates the average altitude $\langle z_{\mathrm{d}}\rangle$.}
    \label{fig:2}
\end{figure}

On Figure \ref{fig:2}(a), we present several examples of  individual droplet dynamics $R(t)$ for $\Delta T=4$K. While  conservation of the  order parameter predicts a regime $R(t)\propto(t_{\mathrm{f}}-t)^{1/3}$, these measurements clearly demonstrate that this scaling is not verified.
Instead,  experimental data show a regime $R(t)\propto(t_{\mathrm{f}}-t)^{\alpha}$, with $\alpha\simeq 0.5-0.6 $. This behavior is confirmed down to  $\Delta T=0.2$K [Fig.~\ref{fig:2}(b)].  
As the decay amplitude varies from drop to drop (Fig. \ref{fig:2}) and is correlated to the mean altitude $\langle z_{\mathrm{d}}\rangle $ (defined as the temporal average of the spanned $z_\mathrm{d}(t)$), 
a natural candidate for an additional physical effect impacting the evaporation could be gravity. Indeed, gravity is well known to affect the statics~\cite{Moldover1979} and dynamics~\cite{ratke1985influence,Bray1994} of phase transitions. We thus develop a theoretical framework to understand the impact of gravity and surface tension on the evaporation dynamics of single droplets. 

\textit{Theory.} 
We consider the free-energy functional 
\begin{equation}
\mathcal{H}[\phi]=\int d\ve[x] \left\{V(\phi)+\frac{\kappa}{2} (\nabla\phi)^2 - \rho_{\text{e}} g z \phi \right\}\label{Hamiltonian},
\end{equation}
where the order parameter $\phi$ is the local micellar volume fraction, $V(\phi)$ is a  double well potential, and $\kappa$ gives the energy cost of local gradients of $\phi$. Also,  $z$  is the vertical coordinate (increasing in the downward direction, with $z=0$ at the meniscus), $\rho_{\text{e}}=\rho_{\mathrm{micelle}}-\rho_{\mathrm{solvent}}$ is the effective mass density of the micelles, and $g$ is the earth gravitational acceleration. 
Next, we assume that the dynamics   satisfies the equations of the (noiseless) model $H$ \cite{hohenberg1977theory}: 
\begin{eqnarray}
	&\partial_t\phi + \ve[u]\cdot\nabla\phi=\lambda\nabla^2 \mu, \ \ \ \ &\mu(\ve[x])=\delta\mathcal{H}/\delta\phi(\ve[x]), \label{ModelH_1}\\
	&\eta \nabla^2 \ve[u]= \nabla p + \phi\nabla\mu, \ \ \ \  &\nabla\cdot\ve[u]=0\label{ModelH_2},
\end{eqnarray}
with $\lambda$  a kinetic coefficient, $\mu$  the  chemical potential, $\eta$  the viscosity and $p$  the pressure. Here, the transport equation (\ref{ModelH_1}) for the conserved order parameter $\phi$ is coupled to Stokes' equation  (\ref{ModelH_2}) describing the flow field $\ve[u]$ in the fluid (assumed to be incompressible) with an additional stress induced by chemical potential gradients. To ensure the conservation of the number of micelles, we impose that $\ve[u]$ and $\nabla\mu$ vanish at the boundaries of the system. 
  
\textit{Sharp interface limit.- } 
We consider the sharp-interface limit, in which $\kappa$ is large enough so that, in each bulk phase $i$, the field $\phi$ is close to its equilibrium value ($\phi\simeq \Phi_i+\overline{\phi_i}$, with small $\overline{\phi_i}$), while $\phi$ varies sharply across  interfaces. 
Here, $\Phi_1$ and $\Phi_2$ are the equilibrium values of $\phi$ on each side of the meniscus. 
In the sharp interface limit, following the arguments of Ref.~\cite{Bray1994} and adding gravity, an effective dynamics is obtained for $\overline{\phi}_i$ with effective boundary conditions at the interface \cite{SM_CriticalShrinking}. 
In the bulk phases, at equilibrium, $\overline{\phi}_i\simeq \rho_{\text{e}} g z/V_i''$, where $V_i''\equiv \partial_\phi^2V\vert_{\Phi_i}$, 
so that, due to gravity, the concentration of micelles varies with the altitude $z$ and differs from its standard value at co-existence equilibrium \cite{giglio1975optical}. 
The dynamics for $\overline{\phi}$ obeys the advection diffusion equation
\begin{equation}
	\partial_t\overline{\phi}_i+\ve[u]\cdot\nabla\overline{\phi}_i= D_i\nabla^2 \overline{\phi}_i, \ \ \ D_i=\lambda V_i''.\label{EqDiff}
	\end{equation}
Next, the boundary conditions at an interface read
\begin{equation}
\overline{\phi}_i =\frac{-\gamma C}{ \Delta\Phi V''_i },	\   \left[\Delta \Phi (\ve[v]-\ve[u])-D_2 \nabla\overline{\phi}_2+D_1\nabla\overline{\phi}_1\right] \hat{\ve[n]}=0,\label{BoubaryConditions}
\end{equation}
where $ \Delta\Phi=\Phi_1-\Phi_2$, $\ve[v]$ is the interface velocity,  $\hat{\ve[n]}$ is the unit  vector normal to the interface (pointing towards  phase $1$) and $C$ is the total interface curvature. The first equation in Eq.~(\ref{BoubaryConditions}) is a form of Gibbs-Thomson relation, while the second one expresses the conservation of the number of micelles.

Further assuming that advective and instationary terms in Eq.~(\ref{EqDiff}) are negligible (so that $\nabla^2\overline{\phi}_i \simeq0$),  these equations can be solved by requiring that the equilibrium solution should be recovered far from the droplet. This leads to
\begin{equation}
\dot{R}= \frac{- \lambda }{R \Delta \Phi}\left(\frac{2\gamma}{R\Delta\Phi}+   z_{\mathrm{d}}  \rho_{\text{e}} g\right), \hspace{0.3cm}
\dot{ z_{\mathrm{d}}}=-\frac{4\rho_{\text{e}}\Delta\Phi g R^2}{15\ \eta}, \label{DynamicsGlobal}
\end{equation}
where $\dot{R}\equiv\partial_tR$. The latter equation corresponds to Hadamard's formula for the sedimentation velocity of a liquid drop.  
Equations (\ref{DynamicsGlobal}) describe the coupling between the evaporation dynamics and gravity. Two effects are at work: (i) first, if the droplet is at an altitude $z_{\mathrm{d}}$, its evaporation dynamics is accelerated due to the increased micelle concentration at this altitude, and second (ii) a droplet moves and experiences different micelle concentrations due to Hadamard's law. 

According to Eq.~(\ref{DynamicsGlobal}), when $R\to0$,  $\dot{z}_{\mathrm{d}}$ vanishes and the final stage of evaporation occurs at constant $z_{\mathrm{d}}$. In this case, integrating the equation for $\dot{R}$ leads to
\begin{equation}
f\left( \frac{z_{\mathrm{d}} R(t)}{L_{\mathrm{c}}^2}\right)=  \frac{\lambda \Delta\Phi (t_{\mathrm{f}}-t)(\rho_{\text{e}} gz_{\mathrm{d}})^3}{\gamma^2},\label{Scaling}
\end{equation}
where $ f(u)=\frac{(u-4) u}{2} +4 \ln \left(\frac{u+2}{2}\right)$ and $L_{\mathrm{c}}=\sqrt{\gamma/(\rho_{\text{e}} g\Delta \Phi)}$  is   the gravitational capillary length. As limiting cases, we retrieve, for small radii,
the law for surface-tension dominated decay
\begin{equation}
R(t)\simeq  \left[\frac{6\lambda \gamma  }{(\Delta\Phi)^2} (t_{\mathrm{f}}-t)\right]^{1/3}, 
\end{equation}
whereas when $R\gg L_{\mathrm{c}}^2/  z_{\mathrm{d}} $, one obtains the gravity-dominated regime
\begin{equation}
R(t)\simeq  \left[a (t_{\mathrm{f}}-t)\right]^{1/2}, \ a=2\frac{\lambda\rho_{\text{e}} gz_{\mathrm{d}}}{\Delta\Phi} \label{GravScaling}.
\end{equation}
This behavior looks similar to the classical $R-$squared laws, but here evaporation is driven by the difference between $\phi$ at the drop altitude and its value at which phase coexistence is allowed.   

\begin{figure}[ht!]
    \includegraphics[width=\columnwidth]{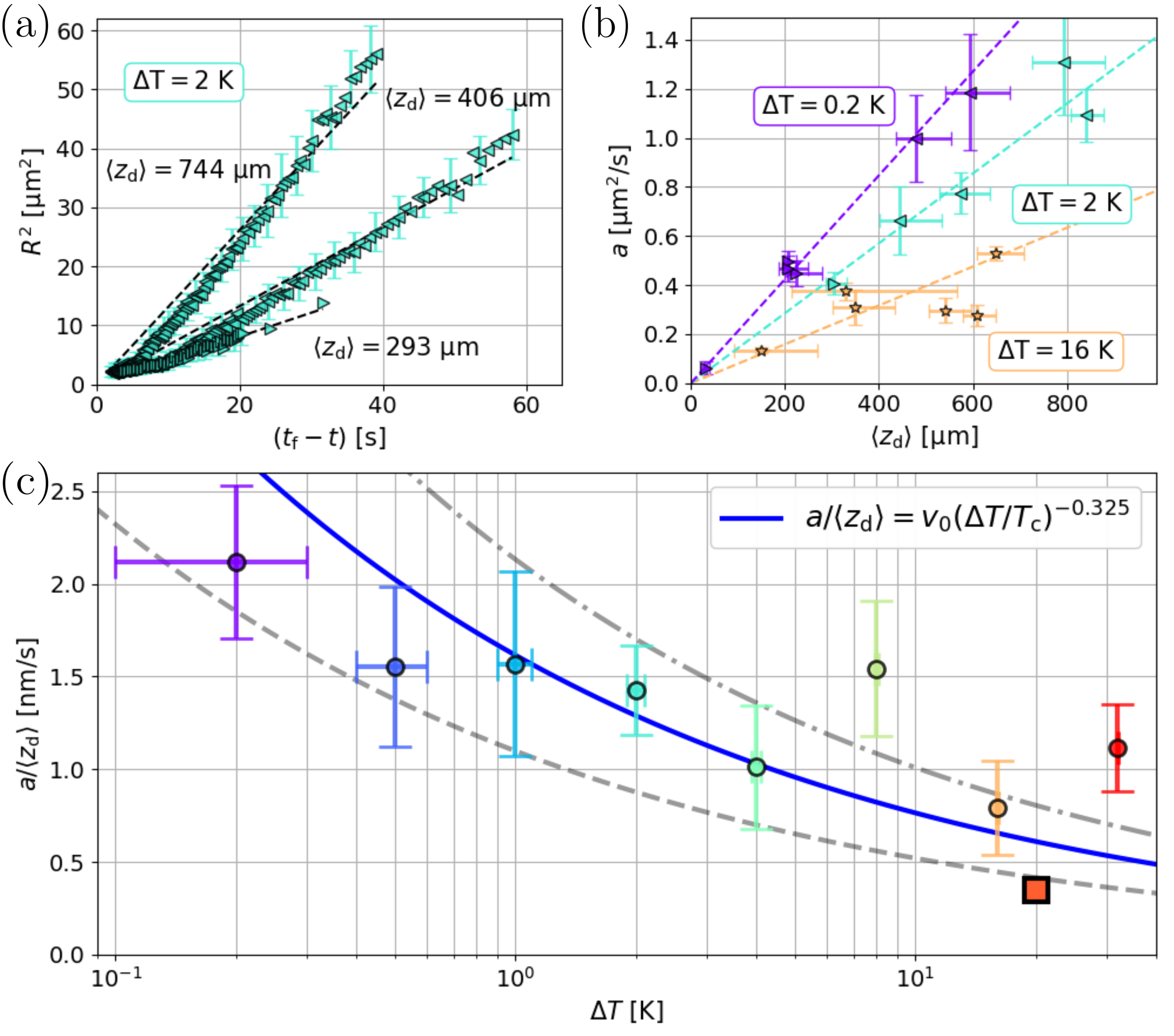}
    \caption{(a) $R^{2}$ as a function of $ t_{\mathrm{f}}-{t}$ for different droplet mean altitude  $\langle z_{\mathrm{d}} \rangle$ for $\Delta T = 2$K. Error bars 
    are shown for one out of four data points for visibility issues. Dashed lines correspond to the linear fit  $R^{2}=a \left( t_{\mathrm{f}}-t \right)$. 
    (b) Fitted slope $a$ as a function of   $\langle z_{\mathrm{d}}\rangle$, for $\Delta T = $0.2, 2 and 16K. The horizontal `error-bars' indicate the range of $z_{\mathrm{d}}$ sampled during the evaporation,  while vertical error bars show the standard deviation on $a$. 
 Dashed lines are linear fits. 
(c) $a/\langle z_{\mathrm{d}} \rangle$ as a function of $\Delta T$. Blue continuous line: best fit with the law  $a/\langle z_{\mathrm{d}} \rangle = v_{\mathrm{0}} \left( \Delta T / T_{\mathrm{C}}\right)^{-0.325}$ with  $v_{\mathrm{0}} = 0.25 \mathrm{nm/s}$. 
Black dashed lines: same law with the coefficients $v_{\mathrm{0}}= 0.17,~ 0.33 \mathrm{nm/s}$. Square symbol: $a/\langle z_{\mathrm{d}}\rangle$ obtained from the theoretical analysis at $\Delta T=20$K (see text). 
} 
    \label{fig:3}
\end{figure}

\textit{Experimental evidence of gravity effects. } On Figure \ref{fig:3}(a), we show experimental curves $R^2$ as a function of time. These curves are compatible with a linear relation  $R^2(t)=a \left( t_{\mathrm{f}}-{t} \right) $. The mean slope $a$ is then reported as a function of  $z_{\mathrm{d}}$ on Figure \ref{fig:3}(b), over a large variation of $\Delta T=0.2-16$K. A linear  dependence of $a$ on $z_{\mathrm{d}}$ is obtained, as predicted in the gravity dominated regime, see Eq.~(\ref{GravScaling}). The experimental values of $a/\langle z_{\mathrm{d}}\rangle$ are reported on Fig.~\ref{fig:3}(c) and show a weak temperature dependency. Theoretically, the diffusion coefficient $D_i=\lambda V_i'' $ is usually believed to scale as $D_i\propto V_i'' $, as demonstrated   for spin exchange dynamics~\cite{kawasaki1966diffusion}. In our case, we conclude that $\lambda$ does not depend on $T$ near the critical point. Then, we expect $a/z_{\mathrm{d}}\propto 1/\Delta\Phi \sim 1/(\Delta T)^\beta$ with $\beta\simeq 0.325$.  Such scaling is compatible with the experiments, if one takes into account the dispersion of the data [Fig.~\ref{fig:3}(c)].  
Thus, observations point towards gravity as being responsible for the acceleration of the evaporation dynamics from the expected exponent $\alpha=1/3$ to the observed one $\alpha\simeq 1/2$.

\begin{figure}[ht!]
    \includegraphics[width=8cm]{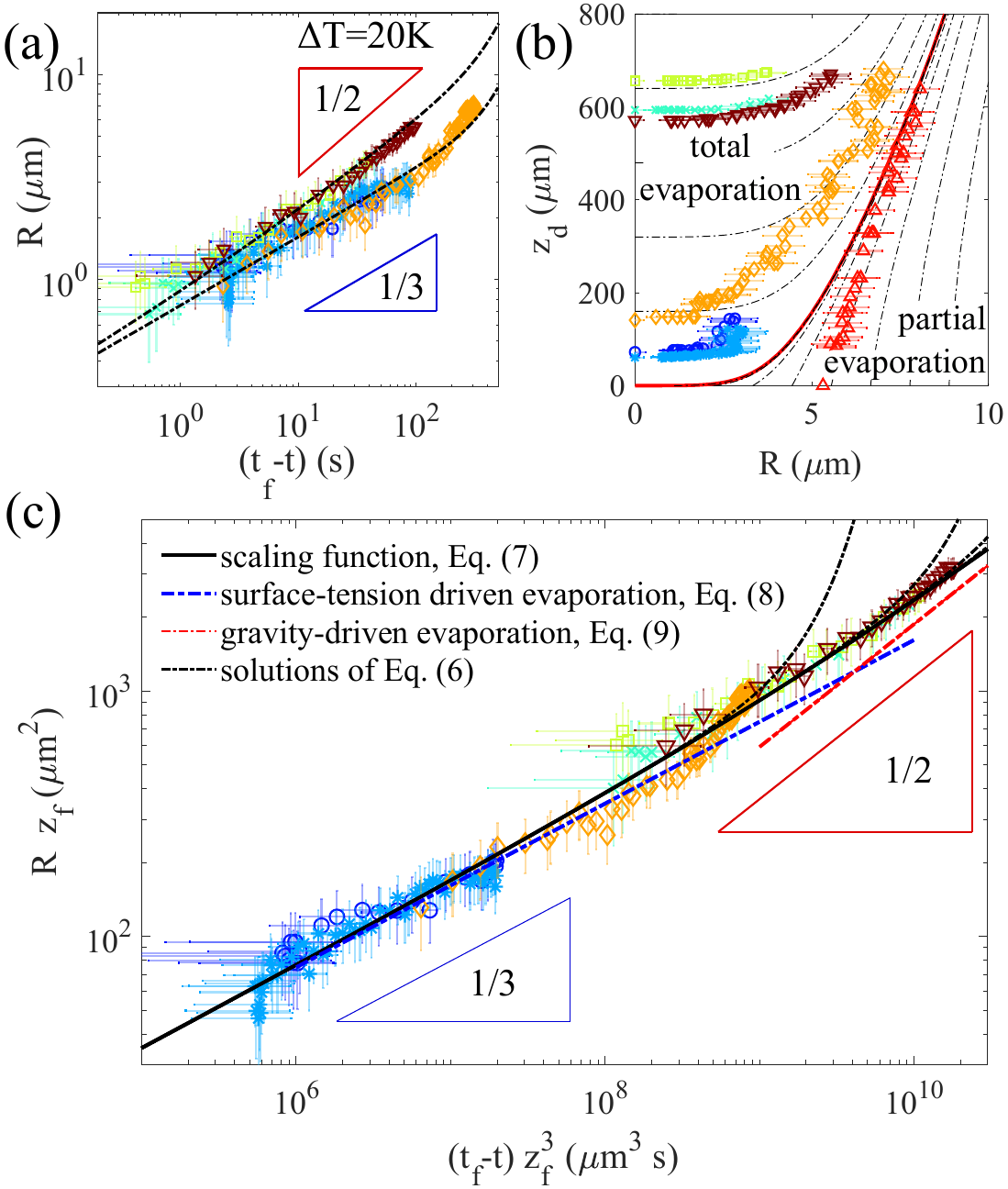}
    \caption{Experimental and theoretical dynamics  for $ \Delta T=20$K. (a) $R$ as a function of $t_\mathrm{f}-t$, (b) same trajectories in the $(R,z_{\mathrm{d}})$ plane and (c) $R(t)$ in rescaled coordinates, with $z_\mathrm{f}\equiv z_\mathrm{d}(t_\mathrm{f})$. Symbols: experimental values for six different drops, with an additional drop in (b) (red triangles). Dash-dotted black lines: solutions of Eq.~(\ref{DynamicsGlobal}). In (c), the continuous black line represents the scaling form (\ref{Scaling}), and dash-dotted blue and red lines indicate respectively the regimes with $\alpha=1/3$ and $\alpha=1/2$. Parameters of the theory: $L_\mathrm{c}=20\mu m$, $\lambda \gamma /(\Delta\Phi)^2=0.07\mu$m$^3$/s, $\rho_{\text{e}}\Delta\Phi g /\eta=0.19\mu$m$^{-1}$s$^{-1}$ \cite{SM_CriticalShrinking}.
    }   
    \label{fig:4}
\end{figure}

\textit{Small drops near the meniscus: surface-tension dominated dynamics - } Nonetheless, the model indicates that the decay exponent  $\alpha=1/3$ should be  retrieved when $R\ll L_{\mathrm{c}}^2/z_{\mathrm{d}}$, \textit{i.e.} when evaporating drops are small, close to the meniscus (where $z_{\mathrm{d}}=0$) and far from criticality (as $L_{\mathrm{c}}$ vanishes at the critical point). 
Since the gradient edge-detection method may be intrinsically limited   for small radii \cite{SM_CriticalShrinking}, we conduct our measurement procedure as follows. Starting from calibrated silica beads of radii $\mathrm{1.15 \mu m}$ and $\mathrm{2.18 \mu m}$ in a water/glycerol mixture to reach refractive index contrasts close to those of our phase-separated system, a 
complementary   image analysis method 
 is developed, in agreement with the previously mentioned method
~\cite{SM_CriticalShrinking}. On Figure \ref{fig:4}(a), we show the decay dynamics at $\Delta T=20$K, relatively far from the critical point. It is observed that some droplets decay  with an exponent $\alpha=1/3$, either over a large temporal window or only at the end of the dynamics, when $t\to t_\mathrm{f}$. 
Fig.~\ref{fig:4}(b) represents the trajectories of the same drops in the $(R,z_{\mathrm{d}})$ plane, where one distinguishes between a region where drops reach the meniscus with a finite size, and a region where drops fully evaporate before reaching the meniscus, in agreement with the flow lines of the dynamical system (\ref{DynamicsGlobal}). 
The comparison with Fig.~\ref{fig:4}(a)  shows that the drops displaying $\alpha=1/3$ during the longest time range are near the meniscus (small $z_{\mathrm{d}}$, see \textit{e.g.} blue symbols), as expected theoretically. We also observe that $\alpha=1/2$ appears when drops are far from the meniscus (large $z_\mathrm{d}$, see \textit{e.g.} brown symbols). To quantify this effect, we rescale our data according to Eq.~(\ref{Scaling}) on Figure \ref{fig:4}(c). The rescaled data sets fall onto a   master curve, in agreement with the expression (\ref{Scaling}), with a fitted parameter $a/z_{\mathrm{d}}$ compatible with other experiments [square on Fig.~ \ref{fig:3}(c)], with a capillary length $L_\mathrm{c}=20\mu $m. 
While within the correct order of magnitude,   this value is 3-4 times smaller than expected. 
This underestimation could be attributed to the use of a simplified mean-field, isothermal and quasi-static model, with an \textit{a priori} assumed model $H$ dynamics, which could describe real dynamics only at the cost of using renormalized coefficients. Nonetheless, even if refined theoretical ingredients may be needed, our model not only predicts the surface-tension ($\alpha=1/3$) and the gravity-driven ($\alpha=1/2$) evaporation regimes and their crossover,  but  also enables a full rescaling of the data set and  a semi-quantitative comparison with experiments, as shown in Fig.~\ref{fig:4}(c).  

\textit{Conclusion.-} With the aim of investigating the surface-tension driven decay at work in  Ostwald-ripening mechanisms, we used an opto-mechanical strategy to  produce the model situation of a single drop of one phase immersed into the other one. Our observations indicated a decay exponent $\alpha\simeq1/2$, which was attributed to a gravity-induced concentration gradient.  However, 
focusing on experimental conditions predicted by a theoretical analysis, we indeed measured  the decay exponent $\alpha\simeq1/3$ characteristic of surface-tension driven decay for conserved systems. Thus, even if the relative variation of   concentration due to gravity is weak  (see estimates in SM \cite{SM_CriticalShrinking}), it can still influence the domain decay, depending of the droplet size $R$, so that the decay exponent usually starts to be 1/2 and may crossover to 1/3 when $R$ decreases enough due to evaporation. The smallness of the concentration gradient is a necessary condition to get the exponent 1/3 but not sufficient. Consequently, our work raises  experimentally and theoretically new insights on the decay component of the evaporation-condensation mechanism, so important in material sciences.
 
\begin{acknowledgments}
The authors thank the  machine and electronic shops of LOMA for their technical contributions. The authors acknowledge financial support from the ANR through the projects FISICS ANR-15-CE30-0015-01 and ComplexEncounters ANR-21-CE30-0020. %
We thank Antoine Aubret for providing calibration beads, Quentin Moreno and Benjamin Sapaly for preliminary experiments, and Justine Roux and Gary Croise for their help in control experiments.  
\end{acknowledgments}

\appendix
\vspace{2cm}
\begin{center}
\textbf{\large{SUPPLEMENTAL MATERIAL}}
\end{center} 

Supporting the main text, we provide  details on:
\begin{itemize}
\item the near-critical properties of the binary liquid used (Section \ref{SectionNearCritical}),
\item the experimental setup (Section \ref{ExpSetUpSection}),
\item the edge detection procedures, data acquisition, calibration and error calculation (Section \ref{SecEdge}),
\item the theoretical analysis (Section \ref{SecTheory}).
\end{itemize}

\section{Near-critical micellar phases of microemulsion}
\label{SectionNearCritical}
The choice of near-critical micellar phases of microemulsion and the properties of the chosen mixture were described in detail previously, for instance in the Supplementary Information of a previous publication on near-critical spreading of droplets \cite{saiseau2022near}. In brief:

1) We used a micellar phase of microemulsion composed of water ($9\%$ wt), oil (toluene, $70\%$ wt), surfactant (sodium-dodecylsulfate, SDS, $4\%$ wt), and alcohol (n-butanol-1, $17\%$ wt), of critical composition at $T_\mathrm{C}=38^\circ$C. As concentrations of water and surfactant are much smaller than that of oil + alcohol, the quaternary mixture organizes at thermodynamic equilibrium as a suspension of surfactant-coated water nanodroplets, the micelles of size $r\simeq 4$nm, dispersed in a toluene continuum; miscible in water and oil, alcohol is present in both water and toluene, and serves as co-surfactant. Consequently, our microemulsion can be considered as a pseudo binary liquid mixture of micelles in toluene. As for any binary mixture, it  exists a line of critical points, which separates the miscible to the phase-separated state, consisting in two micellar phases of different micellar concentrations and the chosen composition is critical at  $T_\mathrm{C}=38^\circ$C.

2) In near-critical conditions, observations obtained from one system are totally transposable to any other belonging to the same Ising universality class, so that one can choose a particular system which is `better adapted' to the dedicated experimental situation without losing generality of the investigation; micellar systems are particularly interesting when looking at supramolecular liquids. When the ratio between water and oil is small, micellar phases of microemulsion belong to the Ising ($d=3, n=1$) universality class, where $d$ and $n$ are respectively the space and the order parameter dimensions, which is representative of isotropic liquids. Close to the critical point, the properties vary as power laws of the temperature shift to the critical temperature with universal exponents related to the Ising class.

3) For the chosen composition (volume fraction $\Phi_C$), the liquid-liquid phase separation (volume fraction $\Phi_{i=1,2}$) occurs by increasing the temperature above the critical temperature $T_\mathrm{C}=38^\circ$C (so-called low critical point) as indicated in the schematic phase diagram shown in Fig.~1 in the main text. 

4) Many critical properties of the microemulsion were already characterized and of particular interest for the present investigation in the two-phase region ($T>T_\mathrm{C}$), we mention \cite{saiseau2022near}:
\begin{enumerate}
\item the correlation length of density fluctuations in the phase separated state, $\xi^-=\xi_0^-[(T-T_\mathrm{C})/T_\mathrm{C}]^{-\nu}$, with $\nu\simeq0.63$ and $\xi_0^-\simeq 2$nm,
\item the micellar volume fraction of the coexisting phases: 
\begin{align}
\Phi_{i=1,2}=\Phi_C+b \left(\frac{T-T_\mathrm{C}}{T_\mathrm{C}}\right)\pm \frac{\Delta \Phi_0}{2}\left(\frac{T-T_\mathrm{C}}{T_\mathrm{C}}\right)^\beta,
\end{align}
with $\beta\simeq 0.325$, the concentration at criticality $\Phi_C=0.11$, the asymmetry parameter $b=1.185$ and the coexistence amplitude $\Delta \Phi_0=0.275$,
\item the density of the two coexisting phases: $\rho_{i}=\rho_{\text{micelles}}\Phi_i+\rho_{\text{solvent}}(1-\Phi_i)$,   $i\in\{1,2\}$, 
with  $\rho_{\text{micelles}}=1040.0$kg/m$^3$  and $\rho_{\text{solvent}}=839.5$kg/m$^3$ the respective densities of the micelles and of the oil continuum,
\item the density contrast between the coexisting phases: $\Delta\rho=\rho_1-\rho_2=\Delta\rho_0[(T-T_\mathrm{C})/T_\mathrm{C}]^{\beta}$, with $\Delta\rho_0=(\rho_{\text{micelles}}-\rho_{\text{solvent}})\Delta\Phi_0=55.1$ kg/m$^3$
(we also found $\Delta\rho_0=53.5$ kg/m$^3$ using the relation $\Delta\rho=(\partial\rho/\partial\Phi)\Delta\Phi$),
\item the interfacial tension between the coexisting phases: $\gamma=\gamma_0 [(T-T_\mathrm{C})/T_\mathrm{C}]^{2\nu}$, 
with $\gamma_0=(5.0\pm0.2) 10^{-5}$N/m,
\item the shear viscosity of the coexisting phases measured empirically:
$\eta=(1.46-0.014\times(T-273\text{K})) (1+2.5\Phi_i)10^{-3}$ Pa.s, 
\item the refractive index contrast between coexisting phases: $\Delta n= 4.62 \times 10^{-2} [(T-T_\mathrm{C})/T_\mathrm{C}]^{-\beta}$, 
\item the osmotic susceptibility in the phase separated state $\chi_T^-$:
\begin{align}
\chi_T^-=\chi_0^- \left(\frac{T_C}{T-T_C}\right)^{3\nu-2\beta}, \label{ValueChi}
\end{align}
where $ 3\nu-2\beta\simeq1.24$ and $ \chi_0^-\simeq1.344\times 10^{-6}\ \text{m}^3.\text{J}^{-1}$ \cite{wunenburger2011fluid}. 
\item while just pertinent for the production of drops by the optical radiation pressure, the optical absorption of the microemulsion is $\alpha_a=(4.8\pm0.5)\ 10^{-4}$cm$^{-1}$,  so that laser heating can be neglected at the used beam powers (smaller and smaller when the critical point is neared).
\end{enumerate}

\section{Experimental setup}
\label{ExpSetUpSection}

\begin{figure}[h!]
\centering
\includegraphics[width=\linewidth,clip]{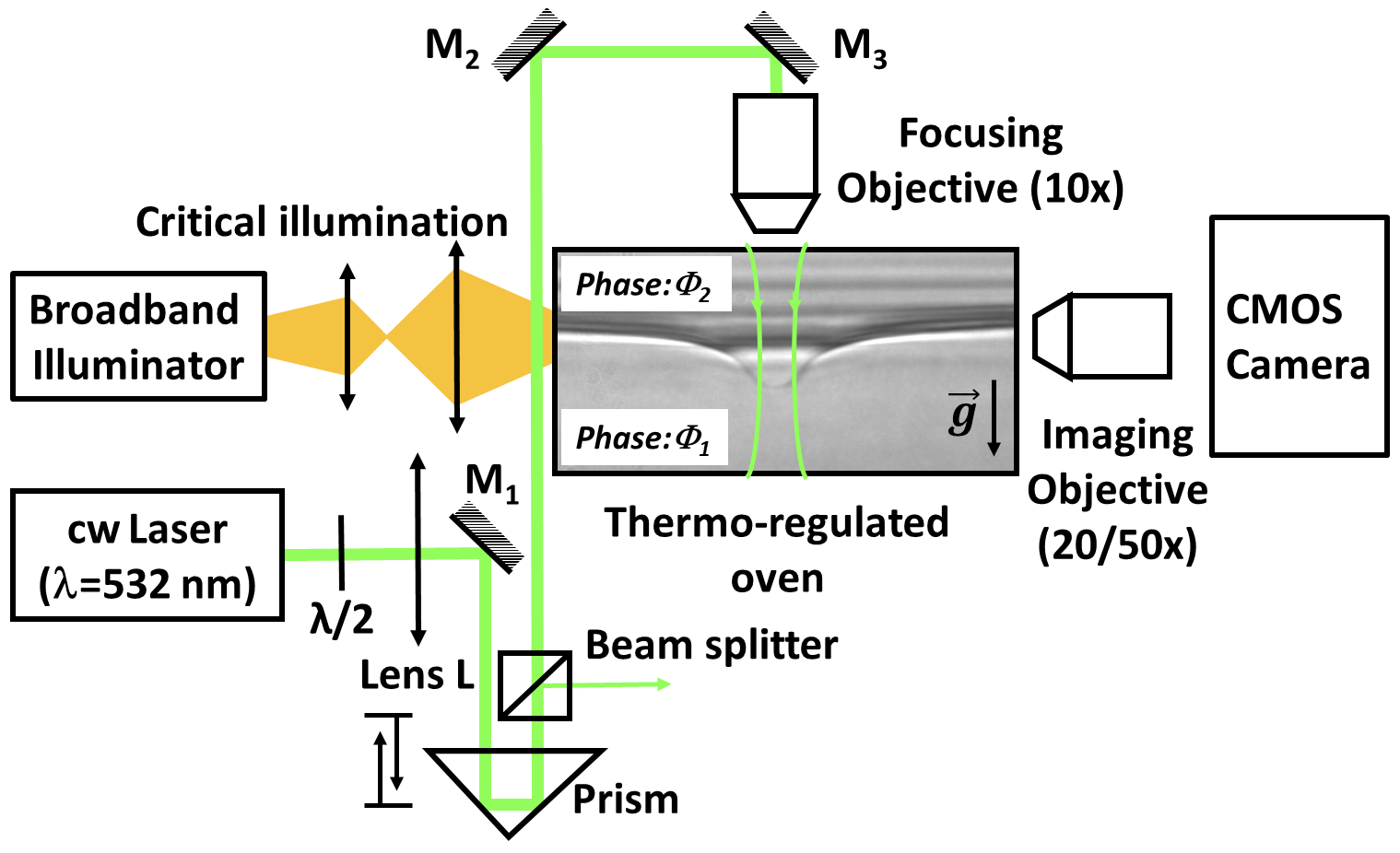}   
\caption{Experimental setup.}
\label{FigExpSetUp}
\end{figure}

The experimental setup was also described several times \cite{wunenburger2006lightDynamics} with regular improvements \cite{Petit2012,girot2019conical,saiseau2022near}. In brief, the schematics on Fig.~\ref{FigExpSetUp} shows that it is composed of three main parts.

1) We focus a continuous laser beam ($\lambda=532$nm, TEM$_{00}$, Coherent Verdi V5) on the meniscus of a phase-separated liquid system, and use the optical radiation pressure to locally deform and eventually destabilize this meniscus to form a liquid jet and droplets by breakup of the phase ($\Phi_2$) into the phase ($\Phi_1$) because the index of refraction of $\Phi_2$ is larger than that of $\Phi_1$; this contactless approach is particularly suitable for the manipulation of tight critical systems. The beam power is controlled using the combination of a half-wave plate and a polarization beam splitter. The beam size at the meniscus (beam waist) is controlled by varying the distance between the lens L ($f=80$cm) and the focusing objective using the movable prism. Even not relevant for the present investigation, where no beam is present, the choice of beam power and waist is essential to drive the meniscus instability in a controlled way \cite{girot2019conical}. 

2) The temperature of the near-critical phase-separated microemulsion is controlled using a brass oven, including resistors driven by a PID controller and a Pt-100 sensor. This oven is also enclosed in a larger box, which includes a thermally-controlled water circulation in order to eliminate slow temperature variation of the lab environment. The temperature stability is locally better than $0.05$K.

3) The illumination part is composed of a high-Intensity Fiber-Coupled Illuminator (OSL2, Thorlabs) and a critical illumination arrangement to focus light at the laser beam location; we made this choice to locally increase lighting, as compared to a K\"ohler illumination scheme, because the turbidity, and thus light scattering, increases drastically close to a critical point. Imaging and frame grabbing are realized using a long working distance objective (due to the bulk volume of the brass oven, Olympus\textsuperscript{\textregistered} LMPLFLN $\times$20 or LMPL $\times$50) and a CMOS high speed camera with a large 4 Mpx sensor (Phantom V EO 340L) to be able to easily follow not only the drop evaporation but also the drop trajectories.

\begin{figure} [!h!]
\includegraphics[width=\columnwidth]{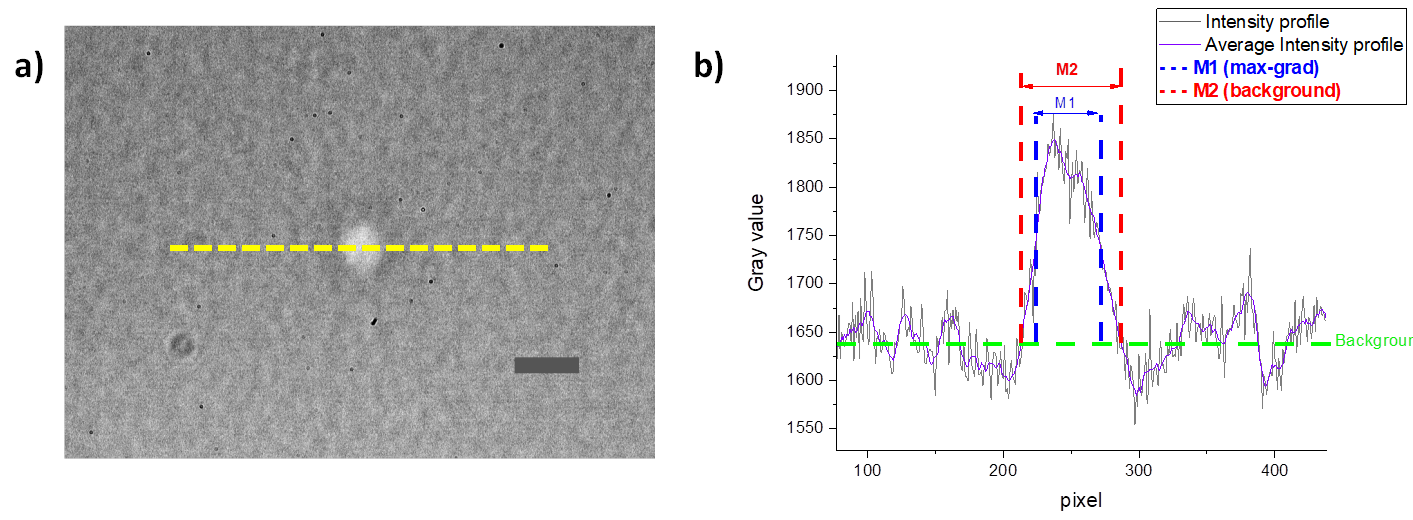}
\caption{a) Image of a calibrated bead ($R= 2.185\pm 0.07$ $\mu m$). The dashed line indicates the line across the object (here a bead) where the intensity profile is analyzed. Scale bar represents 100 pixels.  b) Intensity profile of the  bead shown in a). The intensity profile is first averaged. In the first edge detection method (M$1_\text{max-grad}$), the diameter of the object is determined by the position of the maxima of intensity gradients. In the second edge detection method (M$2_{\mathrm{background}}$), the size of the object is determined when the intensity is equal to the background intensity. }%
\label{edge_detection}%
\end{figure}

\section{Technical details on edge detection procedures, data acquisition, calibration and considerations on error calculation}
\label{SecEdge}

\subsection{Overview}
Here, we present the edge detection procedure used in the main article to measure the diameter of the drops.  Two edge detection procedures have been developed. The first edge detection method - called M$1_{\text{max-grad}}$ method - is based on the position on the maxima of intensity gradients (see in figure~\ref{edge_detection}) and has been described in details in Ref.~\cite{saiseau2022near}. Because this method is computationally non-ambiguous, it can be easily implemented algorithmically, and is robust to global intensity variations~\cite{van2014velocity}. However, as already mentioned in Refs.~\cite{van2014velocity,saiseau2022near}, this method tends to underestimate the drop radius by a systematic additive error called $d_{\mathrm{cal}} $, which can be estimated thanks to the analysis of  volume conservation (see below). The radius of the object is then given by : $R^{\mathrm{exp}}=R_{\mathrm{max-grad}}+d_{\mathrm{cal}} $. However, this systematic error limits intrinsically the detection of very small drops (when $R\rightarrow 0$).

\begin{figure}%
\includegraphics[width=\linewidth]{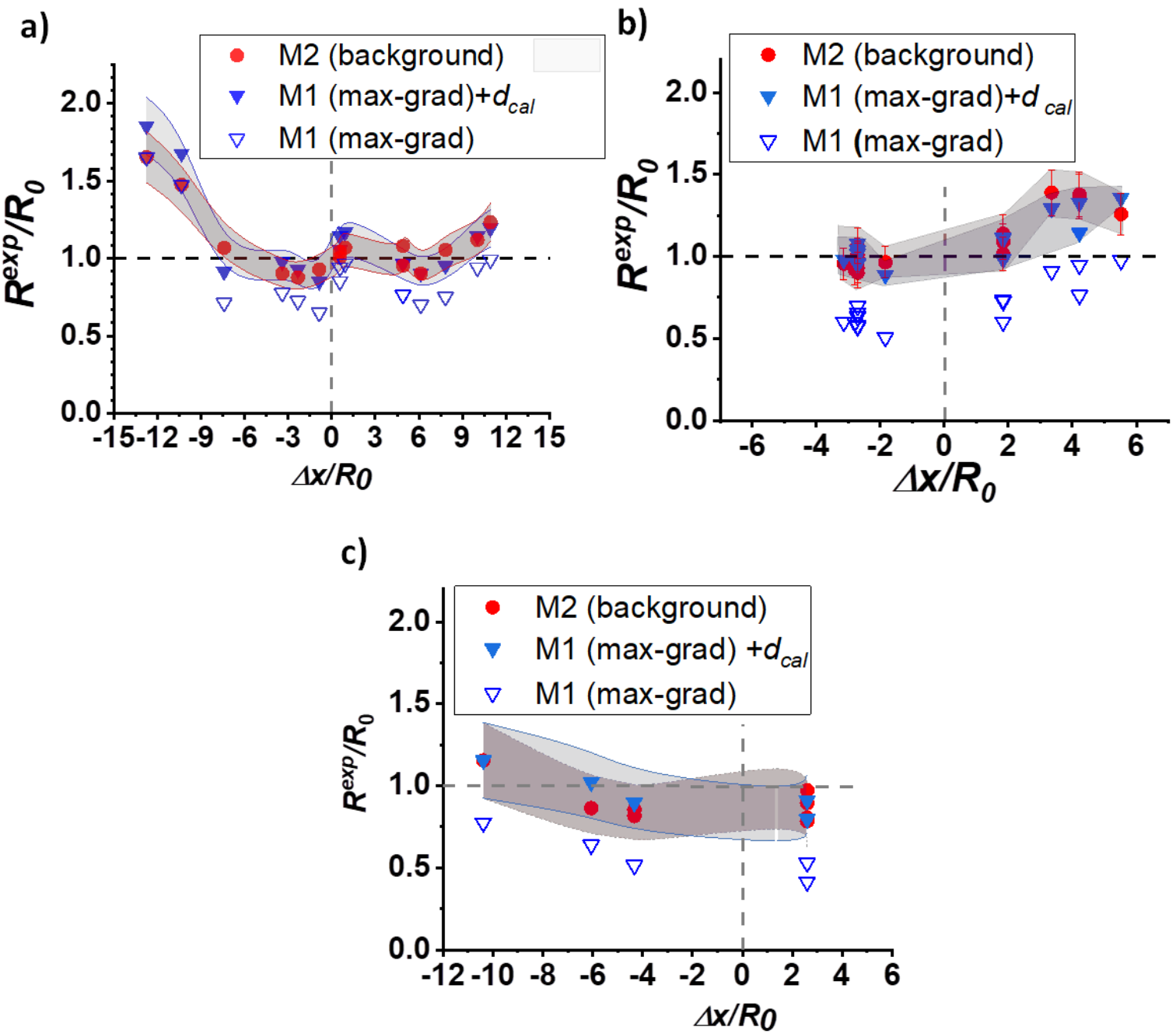}%
\caption{Calibration curves of the measured radius  $R^{\mathrm{exp}}$  renormalized by the calibrated radius  $R_0$ of the bead as a function of the renormalized defocusing distance $\Delta x/ R_0$. Open triangles represent measurements with M$1_\text{max-grad}$ method. Filled triangles represent results with the M$1_\text{max-grad}$ with the correction $d_\mathrm{cal}$.  Filled circles represent results with the  M$2_{\mathrm{background}}$ method.  a) Calibration curve for the 20X objective and $R_0=2.185$ $\mu$m. b) Calibration curve for the 20X objective for  $R_0=1.155$ $\mu$m. c) Calibration curve for the 50X objective for $R_0=1.155$ $\mu m$. Error-bands indicate the uncertainty  on $R_1^{\text{exp}}$ and $R_2^{\text{exp}}$, respectively. }%
\label{calibration_curves}%
\end{figure}

To overcome these limitations, we have   implemented a second edge detection method called hereafter M$2_{\mathrm{background}}$ method. This method is based on the pixel positions where the intensity is equal to the background image intensity (see  figure~\ref{edge_detection}). Importantly, we have validated the size measurements obtained with this  method using calibrated beads at $R=1.15$ $\mu m$ and $R= 2.185$ $\mu m$. 
Finally, as shown in figure~\ref{calibration_curves}, we have checked that the two detection methods (the M$1_{\text{max-grad}}$ corrected with $d_{\mathrm{cal}} $ and the M$2_{\mathrm{background}}$ methods) are compatible with calibrated values (within $20 \%$) over a wide range of defocusing distances. Below we give the full details of these edge detection procedures. 

\subsection{ First edge detection method based on maximal intensity gradient: M$1_\text{max-grad}$ method }
The first edge detection method used in the main text  has been developed by Bertollini \textit{et al.}~\cite{bertollini1985image}  to measure aerosols droplets distributions and has been described in details in Ref.~\cite{saiseau2022near}. It is based on the positions of the maxima of the intensity gradients in the image.\\

\paragraph{Radius and position determination\\} 

\begin{figure}[!h!]
\includegraphics[width=\columnwidth]{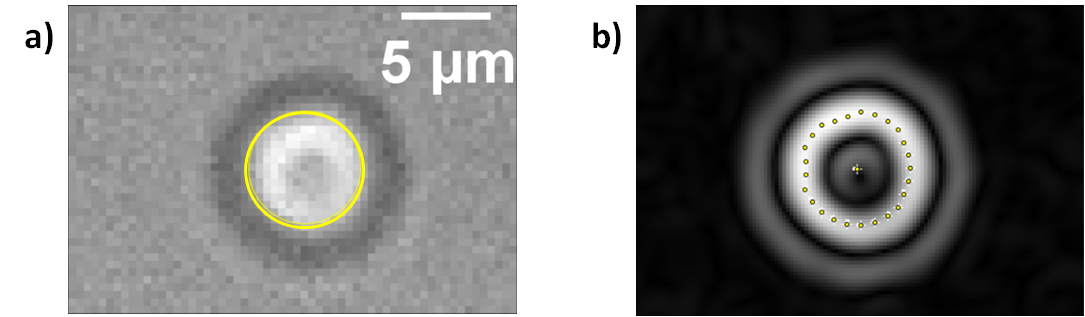}%
\caption{a) Microscopy image of a near-critical microemulsion drop at $\Delta T=4$ K. In yellow is shown the reconstructed circle with a radius $R=3.36 \mu m$ obtained from the intensity gradient maxima and the calibrated distance. b) Intensity gradients obtained using a Sobel kernel method of image shown in a). The $N=25$ intensity gradients maxima obtained for uniformly distributed angular directions are shown as well as their centroid. }%
\label{fig_detection}%
\end{figure}

Here, we describe the algorithmic procedure  to obtain the droplet position and radius. 
For each image:
\begin{itemize}
	\item an image of the background made after the droplet is completely evaporated is subtracted and a Gaussian filter is applied using the microscope objective resolution for the kernel size ($\sim 1$ $\mu m$),
	\item a first rough estimate of $R$ is obtained using an automated circle localization algorithm from the imageJ plug-in ``Analyze Particles'',
	\item from the circle center, the positions of the intensity gradient maxima  are obtained using a Sobel kernel for $N=25$ directions with uniformly distributed angles [Fig.~\ref{fig_detection}(b)],
	\item the drop center coordinates $(y_{\mathrm{d}},z_{\mathrm{d}})= (\langle y_i \rangle_N,\langle z_i \rangle_N)$ are then recalculated from these positions and used to follow drop motion. The drop radius is then obtained from $R_{\mathrm{max-grad}}=\langle [(y_i-y_{\mathrm{d}})^2+(z_i-z_{\mathrm{d}})^2]^{\frac{3}{2}} \rangle_N^{1/3}$ .
\end{itemize}
\vspace{0.5cm}

\paragraph{Volume conservation calibration\\}

\begin{figure}[h!]%
\includegraphics[width=\columnwidth]{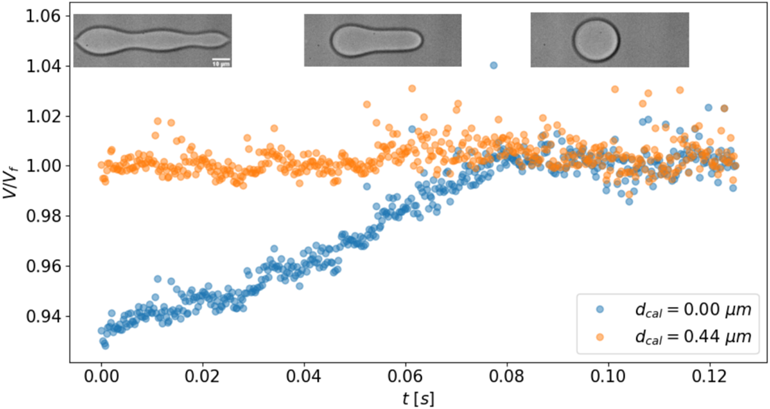}%
\caption{Volume $V$ of a liquid ligament, normalized by its final value, obtained following the method of Ref.~\cite{van2014velocity}, with (orange) and without (blue) taking into account a calibrated distance $d_{\text{cal}}=0.44$ $\mu m$.  Here,  $\Delta T=20$K. Snapshots: evolution of the shape of the liquid filament. }%
\label{volume_cons}%
\end{figure}

Using the intensity gradient maxima, the droplet radius is likely to be underestimated. To account for such systematic error, following the method developed by van der Bos et al.~\cite{van2014velocity}, a calibrated distance $d_{\mathrm{cal}}$ is estimated by insuring volume conservation for strongly elongated liquid object that relaxes towards a spherical shape, see Fig.~\ref{volume_cons}. Indeed, the characteristic time for the relaxation dynamics ($\tau \approx  0.1 $ $s$) is much shorter than the evaporation time $(t_\mathrm{f} \approx 100$ s), so that the volume can be considered as constant during the relaxation. In figure~\ref{volume_cons}, we show the evolution of the volume of the liquid ligament during its relaxation to a sphere, with and without taking into account a calibrated distance $d_{\mathrm{cal}}=0.44\mu$m, which is optimised to ensure volume conservation. This calculation of $d_{\mathrm{cal}}$ has been systematized for various $\Delta T=20$, $10$, $4$ and $2 K$ and drop sizes (ranging between $R=2.5\mu$  m  and $R=6.3 \mu$ m). The best calibrated distance is found to be $d_{\mathrm{cal}} =0.44 \pm 0.28 $ $\mu m$  with no specific dependency on $\Delta T$ or $R$. 

We call $R^{\mathrm{exp}}_{1}$ the measured radius obtained with this first edge detection method, which is thus given by: 
\begin{align}
R^{\mathrm{exp}}_{1}=  R_{\mathrm{max-grad}}+d_{\mathrm{cal}}^0 \pm e_{R},
\end{align}
with $d_{\mathrm{cal}}^0 =0.44 $ $\mu m$.
The uncertainty $e_R$ is then given by $e_R=[\sigma_R^2+ \delta_p^2/12+e_d^2]^{1/2}$, with $\sigma_R$   the standard deviation of the distribution of maxima positions from the droplet center, $\delta_p$ the image resolution, and  $e_d$ the error on the calibration distance $d_\text{cal}$. 

 As shown in figure~\ref{calibration_curves}, the measured radius by the M$1_{\text{max-grad}}$ method alone underestimates the size of the bead. However, once the calibrated distance $d_{\mathrm{cal}}$ is taken into account, the measured radius $R^{\mathrm{exp}}_{1}$ is compatible with the calibrated values for different imaging objectives. As already mentioned, this added calibration distance $d_{\mathrm{cal}}$ is negligible for large drops but it limits intrinsically the detection of very small drops. To overcome this limitation, we develop a second image edge detection technique.

\subsection{Second edge detection technique : M$2_{\text{background}}$ method}
We have implemented another edge detection method, based on the image analysis of calibrated beads. We use two silica calibrated beads (Microparticle GmBH) with diameters $D_1=2.31 \pm 0.05$ $\mu m$ and $D_2=4.37 \pm 0.14$ $\mu m$, where the uncertainty indicates the standard deviation of the distribution. 
\vspace{0.5cm}

\paragraph{Preparation of solutions of calibrated beads in water-glycerol mixtures.\\}
We dilute  the initial bead solution into water-glycerol mixtures. First, we prepare water-glycerol solutions from $57\%$ to $85\%$ in glycerol. For each solution, we measure the refractive index with an Abbe refractometer. 
Secondly, we determine the refractive index of calibrated beads by phase contrast microscopy as described in Ref.~\cite{barer1954refractometry}. Briefly, we image calibrated beads into water-glycerol solutions with different refractive indexes. When the contrast of the bead is lost in phase contrast microscopy, it means that the refractive index of the solution matches the refractive index of the object. We find $n_{\text{bead}}=1.428\pm0.001$ in agreement with the commercial value $n_{\text{bead}}=1.42\pm0.01$.
 To match the optical experimental conditions, we dilute the calibrated beads into a $57 \%$ glycerol-water mixture ($n=1.408$) which gives a refractive index contrast between the bead and the solution $\Delta n \approx  0.02$, which is close to the index contrast between separated phases in the micro-emulsion at $\Delta T= 20K$. 

\vspace{0.5cm}
\paragraph{Image analysis for calibrated beads and validation of the $\mathrm{M}2_{\mathrm{background}}$ method.\\}
We image the calibrated beads immersed in water-glycerol mixtures by varying the focus of the imaging objective with a translation stage in order to get the intensity profile for each defocusing position $\Delta x$ (see figure~\ref{FigExpSetUp} and ~\ref{calibration_curves}). In the M$2_{\text{background}}$ method, the diameter of the object is measured when the intensity is equal to the  background intensity (see figure~\ref{edge_detection}). Calibration results are shown in figure~\ref{calibration_curves} for two calibrated beads $R_0=1.155$ $\mu m$ and $R_0=2.185$ $\mu m$ and various imaging objectives in the setup.  The experimental   radii $R^{\mathrm{exp}}_{2}$ are compatible with the calibrated radii within $20\%$ over a wide defocusing range ($-3 \leq\Delta x/R_0 \leq 3$). It is also compatible with the first method. This allows us to use the M$2_{\text{background}}$ method down to $R=1.155$ $\mu m$ unambiguously. 
 
\vspace{0.5cm}

\section{Theory}
\label{SecTheory}
\subsection{Field profile at equilibrium}

Here we derive the effective equations of the dynamics in the sharp interface limit. Our approach consists in adapting the arguments  presented in Ref.~\cite{Bray1994} to our situation. First, let us describe the profile of $\phi$ inside the meniscus at equilibrium. The chemical potential is given by
\begin{align}
\mu=\frac{\delta \mathcal{H}}{\delta\phi}=-\kappa \nabla^2 \phi + V'( \phi)-\rho_{\mathrm{e}} g z,
\end{align}
where the prime denotes the differentiation with respect to $\phi$. At equilibrium, $\mu$ is uniform, $\mu(\ve[x])=\mu_e$. Let us call $\phi^*(z)$ the profile of $\phi$ across the meniscus. Since  the meniscus is located in the vicinity of $z=0$, we   neglect the gravity term  in the equation $\mu=\mu_e$, which reads
\begin{align}
\kappa \frac{\partial^2 \phi^*}{\partial z^2}-V'(\phi^*) =-\mu_e \label{80193}.
\end{align} 
This equation is  the same as the force-balance equation for a classical point-like particle moving in the effective ``potential'' $-V+\mu_e\phi^*$, with $z$ the ``time'', $\kappa$ the ``mass'' and $\phi^*$ the ``position''. Consequently, the fact that Eq.~(\ref{80193}) must admit solutions that converge to constant values at $z\rightarrow\pm \infty$ imposes that this effective potential has two extrema at $\phi=\Phi_1$ and $\Phi_2$, and takes the same value at these extrema, so that
\begin{align}
	V'(\Phi_1)=V'(\Phi_2)=\mu_e=	\frac{V(\Phi_1)-V(\Phi_2)}{\Phi_1-\Phi_2}.
\end{align}
Hence,  the equilibrium values  $\Phi_1$ and $\Phi_2$ on each side of the interface are obtained by using the method of common tangents. We now define $U(\phi)=V(\phi) -\mu_e \phi$, which has two minima of equal depth at $\Phi_1$ and $\Phi_2$. The interfacial tension $\gamma$, that is the excess energy per unit area of the interface can be calculated as
\begin{align}
\gamma=\int_{-\infty}^{\infty}dz \left\{\frac{\kappa}{2} (\partial_z \phi^*)^2 +U(\phi^*(z))-U(\Phi_i)\right\}. \label{DefGamma}
\end{align}
Far from the interfaces, the field $\phi$ varies due to gravity. We recall that $\overline{\phi}_i(\ve[x])=\phi(\ve[x])-\Phi_i$. Expanding the equation $\mu=\mu_e$ at leading order in $\overline{\phi}_i$, we obtain
\begin{align}
\overline{\phi}_i=\rho_{\mathrm{e}} g z / V''_i, \label{EqPhi}
\end{align}
which describes the increase of micelles concentrations due to gravity when the coordinate $z$ increases. 

\subsection{Dynamics in the sharp interface limit}

In the presence of curved interfaces, or interfaces that are not at $z=0$, the system is no longer stationary. When interfaces are thin, the system can be divided into domains where $\phi$ is close to its equilibrium value. Writing again $\overline{\phi}_i(\ve[x])=\phi(\ve[x])-\Phi_i$, the evolution of $\overline{\phi}_i$ follows from the expansion of Eq.~(2) at leading order in $\overline{\phi}_i$, and is described by
\begin{align}
\partial_t\overline{\phi}_i + \ve[u]\cdot\nabla \overline{\phi}_i=D_i \nabla^2 \overline{\phi}_i \label{94014532}. 
\end{align}
To obtain effective boundary conditions at the interface, we consider the chemical potential $\mu$ at the interface located at the altitude $z=z_s$:
\begin{align}
\mu-\mu_e=\ &U'(\phi)-\kappa \nabla^2 \phi-\rho_{\mathrm{e}} g z_s, \nonumber\\
\simeq\ & U'(\phi)-\kappa \ \partial_s^2 \phi-\kappa C \partial_s \phi-\rho_{\mathrm{e}} g z_s\label{94103042},
\end{align}
where we have used differential geometry formulas as described in Ref.~\cite{onuki2002phase,Bray1994}. Here, $s$ is the normal coordinate  along the unit vector $\hat{\ve[n]}$ normal to the interface (pointing towards the phase $\Phi_1$ by convention), and $C$ the interface total curvature (when the orientation of the interface is $\hat{\ve[n]}$). Approximating $\phi(s)$ by the equilibrium interface profile $\phi^*$ [defined by Eq. (\ref{80193})], integrating Eq.~(\ref{94103042}) over $s$ after multiplication by $\partial_s\phi^*$ leads to
\begin{align}
\mu-\mu_e+\rho_{\mathrm{e}} g  z_s=- \frac{\gamma C}{ \Delta \Phi} \label{9401}, 
\end{align}
where we have used Eq.~(\ref{DefGamma}). In the bulk phases, the chemical potential is approximated by:
\begin{align}
\mu-\mu_e\simeq V''_i \bar{\phi}_i-\rho_{\mathrm{e}} g z  \label{764110}. 
\end{align}
Comparing Eqs.~(\ref{764110}) and  (\ref{9401}), we see that the boundary condition for the field $\phi$ at a position $\ve[r]_s$ of the interface is
\begin{align}
	V''_i \overline{\phi}_i\vert_{\ve[r]_s}=- \frac{ \gamma C}{ \Delta\Phi},  \label{GTEq}
\end{align}
which is the Gibbs-Thomson equation in our system. Another boundary condition comes from the conservation of the number of micelles at the interface, which writes 
\begin{align}
\Delta \Phi \ (\ve[v]-\ve[u])\cdot\hat{\ve[n]}=[D_2 (\nabla\overline{\phi}_2)_{\ve[r]_s} -  D_1(\nabla\overline{\phi}_1)_{\ve[r]_s}]\cdot \hat{\ve[n]}, \label{EqCons}
\end{align}
where $\ve[v]$ is the velocity of the interface. 

Last, since $\nabla\mu$ is small, Stokes' equation can be approximated by
\begin{align}
\eta \nabla^2 \ve[u]&\simeq \nabla p \ + [\Phi_1\theta_1(\ve[x])+\Phi_2 \theta_2(\ve[x])] \nabla \mu \label{7492905},
\end{align}
where $\theta_i(\ve[x])$ is the function that is unity in the phase $i$ and zero elsewhere.   
This equation can also be written
\begin{align}
\eta \nabla^2 \ve[u]&\simeq \nabla \tilde{p} \ + \gamma C \nabla \theta_1(\ve[x])\nonumber\\
&+ [\Phi_1\theta_1(\ve[x])+\Phi_2 \theta_2(\ve[x])] (-\rho_{\mathrm{e}} g \hat{\ve[e]}_z),
 \label{89421}
\end{align}
where $\tilde{p}=p+(\Phi_1\theta_1+\Phi_2\theta_2)V_i''\overline{\phi}_i $ is a renormalized pressure. Since $\nabla\theta_1$ vanishes everywhere except at the interfaces, the term proportional to $\gamma$ in the above equation induces a discontinuity of the normal stress of the fluid, while the last term is the usual gravity acceleration. Equation  (\ref{89421}) can thus be identified as the equation for the flow of a fluid presenting phases of different densities $\rho_{\mathrm{e}}\Phi_i$, with surface tension $\gamma$. 

Let us finally note that the same equations would be found  by considering a more general dynamics: 
\begin{align}
\partial_t\phi+\ve[u]\cdot\nabla \phi=  \nabla [\lambda(\phi) \nabla\mu] 
\end{align}
which includes Dean's equation ($\lambda(\phi)=\tilde{\lambda}\phi$) \cite{Dean1996}  or previous equations used to study gravity $\lambda(\phi)=\lambda_0-\lambda_1(\phi-\Phi_C)^2$ \cite{puri1994phase,yeung1992phase}. However, at our level of approximations, where one assumes that $\phi$ is close to $\Phi_i$ in each phase, modifying the dynamics would only lead to a renormalization of $D_i=\lambda(\Phi_i) V_i''$. 

\subsection{Dynamics of a spherical domain}

Let us now derive the equations of the dynamics for a domain  $\Phi_2$ with spherical shape immersed inside the phase $\Phi_1$, of radius   $R$ and altitude $z_{\mathrm{d}}$. We assume $z_{\mathrm{d}}\gg R$, which is always satisfied in the present experiments. We solve the above equations (\ref{89421}), (\ref{EqCons}), (\ref{GTEq}) and (\ref{94014532}), requiring that far from the drop (``at infinity'') the equilibrium solution  (\ref{EqPhi}) is retrieved, with a vanishing flow $\ve[u]$. 

First, the flow is that of a spherical drop immersed in a fluid with different density. Its expression  is well-known~\cite{happelBrennerBook1983}. In particular, the radial flow at the interface reads 
\begin{align}
&u_r = \ve[u]\cdot\hat{\ve[e]}_r= - \frac{4 R^2  \rho_{\mathrm{e}} g \Delta \Phi}{15\eta}   \cos{\theta}  \label{8110},
\end{align}
where $\hat{\ve[e]}_r$ is the normal vector pointing to the exterior of the drop,  and $\theta$ is the angle between $\hat{\ve[e]}_r$ and $\hat{\ve[e]}_z$.  

 For the fields $\overline{\phi}_i$, we assume that the dynamics of $R$ is slow enough so that a quasi-stationary limit can be reached and the term $\partial_t\overline{\phi}_i$ can be neglected. We also assume that the Peclet number is small, so that Eq.~(\ref{94014532}) becomes $\nabla^2\overline{\phi}_i=0$. Its solution, which satisfies the boundary conditions is
\begin{align}
\overline{\phi}_1=&\frac{-2 \gamma}{\Delta \Phi V''_1 r }+\frac{\rho_{\mathrm{e}} g z_{\mathrm{d}}}{V''_1}\left(1-\frac{R}{r}\right)+\frac{\rho_{\mathrm{e}} g}{V''_1}(z-z_{\mathrm{d}})\nonumber\\
&-\frac{\rho_{\mathrm{e}} g R^3(z-z_{\mathrm{d}})}{r^3 V''_1},
\end{align}
with $r$ the distance to the center of the drop. The field $\overline{\phi}_2$ inside the drop is uniform.  In particular, the diffusive flux reads
\begin{align}
j_r=(-D_1\nabla \overline{\phi}_1\cdot\hat{\ve[e]}_r)_{r=R} =  \frac{-2 \gamma \lambda }{\Delta \Phi  R^2 }-\frac{\lambda\rho_{\mathrm{e}} g z_{\mathrm{d}}}{R}-3\lambda\rho_{\mathrm{e}} g \cos\theta.
\end{align}

Next, the radial velocity of the interface results from the variation of the drop's radius  and the translation of its center-of-mass:
\begin{align}
v_r=\ve[v]\cdot \hat{\ve[e]}_r=(\dot{R}\ \hat{\ve[e]}_r +\dot{z}_\mathrm{d} \ \hat{\ve[e]}_z) \cdot\hat{\ve[e]}_r= (\dot{R}+\dot{z}_\mathrm{d}\cos\theta)  . 
\end{align}
Using the conservation equation (\ref{EqCons}), we obtain
\begin{align}
\Delta\Phi& \left[\dot{R}+\dot{z}_\mathrm{d}\cos\theta + \frac{4 R^2  \rho_{\mathrm{e}} g \Delta \Phi}{15\eta}   \cos{\theta}  \right] =\nonumber\\
&  \frac{-2 \gamma \lambda }{\Delta \Phi  R^2 }-\frac{\lambda\rho_{\mathrm{e}} g z_{\mathrm{d}}}{R}-3\lambda\rho_{\mathrm{e}} g \cos\theta.
\end{align}
This equation leads to
\begin{align}
&\dot{R}=- \frac{\lambda }{R \Delta \Phi}\left(\frac{2\gamma}{R\Delta\Phi}+   z_{\mathrm{d}}  \rho_{\text{e}} g\right), \label{DynamicsRadius}\\
&\dot{z}_{\mathrm{d}}=-\frac{4\rho_{\text{e}}\Delta\Phi g R^2}{15\ \eta}-\frac{3\lambda\rho_{\mathrm{e}} g}{\Delta\Phi} \label{DynamicsGlobal}.
\end{align}
The first term in the equation for $\dot{z}_{\mathrm{d}}$ is Hadamard's law. The second term in Eq.~(\ref{DynamicsGlobal}) comes from a different mechanism of motion, where phase $\Phi_1$ is converted into phase $\Phi_2$ above the drop, while phase $\Phi_2$ is converted into phase $\Phi_1$ below the drop. This term turns out to be negligible and will be discarded it in the analysis.

\subsection{Regimes of the dynamics}

Defining $\tilde{R}=R/L_R$, $\tilde{z}_\mathrm{d}= z_{\mathrm{d}}/L_Z$ and $\tilde{t}=(t_\mathrm{f}-t)/\tau$, we obtain the set of equations
\begin{align}
	&\partial_{\tilde{t}}\tilde{R}=\frac{2}{\tilde{R}^2}\left(1+\frac{\tilde{z}_\mathrm{d} \tilde{R}}{2}\right)\label{EqRtilde} ,\\
&\partial_{\tilde{t}}\tilde{z}_\mathrm{d}=\frac{4}{15}\tilde{R}^2,
\end{align}
where 
\begin{align}
\tau = \frac{1}{\rho_{\mathrm{e}} g}\sqrt{\frac{\eta}{\lambda}}, \  
L_R=\left[\frac{\gamma \sqrt{\eta \lambda}}{(\Delta\Phi)^2  g \rho_{\mathrm{e}}}\right]^{1/3}, \ 
L_Z=\frac{L_{\mathrm{c}}^2}{L_R}.
\end{align}
With this choice of rescaled variables, the dynamics does not depend on any parameter. Note that we have inverted the time, so that $\tilde{t}=0$ corresponds to $\tilde{R}=0$, and the spherical domain grows when $\tilde{t}$ increases; $\tilde{z}_\mathrm{d}(0)$ is the dimensionless final altitude of the drop. Different regimes can be obtained as follows. First, at the late stage of the dynamics ($\tilde{t}\to0$), we may consider that the dynamics occurs at constant $\tilde{z}_\mathrm{d}=\tilde{z}_\mathrm{d}(0)\equiv\tilde{z}_\mathrm{f}$, so that Eq.~(\ref{EqRtilde}) leads to 
\begin{align}
&f\left( \tilde{z}_\mathrm{f} \tilde{R}\right)=   \tilde{t}\ \tilde{z}_\mathrm{f}^3 , \\
& f(u)= \int_0^u dy \ \frac{ y^2}{(2+y)}=\frac{(u-4) u}{2} +4 \ln \frac{u+2}{2} .\label{Scaling}
\end{align}
Using this formula and the asymptotics of $f$ for large and small $u$, we find that when $\tilde{R}\tilde{z}_\mathrm{f}\ll1$, $\tilde{R}=(6\tilde{t})^{1/3}$, whereas when $\tilde{R}\tilde{z}_\mathrm{f}\gg1$, we find a $R$-squared laws $\tilde{R}= \sqrt{2  \tilde{z}_\mathrm{f} \tilde{t}}$. 

Next, for very large $\tilde{t}$, drops are so big that one can neglect the term which does not contain $\tilde{z}_\mathrm{d}$ in Eq.~(\ref{EqRtilde}), leading to
 \begin{align}
	&\partial_{\tilde{t}}\tilde{R}=\frac{\tilde{z}_\mathrm{d}}{\tilde{R}}, &\partial_{\tilde{t}}\tilde{z}_\mathrm{d}=\frac{4}{15}\tilde{R}^2,
\end{align}
which predicts exponential solutions 
\begin{align}
&\tilde{R}\propto e^{\tilde{t}\sqrt{2/15}}. \label{ExpSol}
\end{align}
To summarize, for $\tilde{t}\gg1$ one obtains exponential solutions (\ref{ExpSol}), while for $\tilde{t}\ll 1$, the law (\ref{Scaling}) holds, which exhibits an $\alpha=1/3$ regime, and possibly an $\alpha=1/2$ regime (if $\tilde{z}_\mathrm{f}$ is large enough so that the $R-$squared law is reached for $\tilde{t}\ll1$, before one switches to the exponential regime.

\subsection{Fit to experiments}

For $\Delta T=20$K, the choice  $\tau\simeq 174$ s, $L_R\simeq2.3\ \mu$m and $L_Z=174\ \mu $m gives a good agreement between the experimental and theoretical trajectories in the $({z}_\mathrm{d},R)$ and the $(R,t)$ representations [Fig.~4 in the main text]. Of note, with these parameters, the coefficient in Hadamard's law reads 
\begin{align}
\frac{\vert\dot{z}_\mathrm{d}\vert}{R^2}=\frac{4 L_Z }{15\tau L_R^2}=0.050 \ \mu \text{m}^{-1}\ \text{s}^{-1} \label{alpheFound},
\end{align}
whereas this coefficient can also be estimated as
\begin{align}
\frac{\vert\dot{z}_\mathrm{d}\vert}{R^2}=\frac{4\rho_{\text{e}}\Delta\Phi g  }{15\ \eta}\simeq 0.060 \ \mu \text{m}^{-1}\ \text{s}^{-1} \label{alphaExp},
\end{align}
where we have used the values given in Section \ref{SectionNearCritical}. The difference between the value (\ref{alpheFound}) and its expected value (\ref{alphaExp}) is thus of the order of $20\%$. 
On the other hand, as mentioned in the main text, $L_\mathrm{c}=\sqrt{L_RL_Z}=20\mu $m, whereas we expect $L_\mathrm{c}\simeq 80\mu$m. This discrepancy could come from (i) the use of an \textit{a priori} assumed  isothermal model $H$ dynamics, without noise, and/or, (ii) possible large scale weak amplitude thermal gradients within the cell containing the microemulsion, due to the presence of four tiny openings within the brass oven, two for the laser beam crossing and two perpendicular ones for lighting and imaging the drop dynamics.

\subsection{Discussion of the approximations}

To discuss the validity of the approximations, we will consider the specific case of a $\phi^4$ potential, $U(\phi)=-\frac{r}{2}(\phi-\phi_0)^2+\frac{u}{4} (\phi-\phi_0)^4$
for which the interface profile can be calculated: 
\begin{align}
\phi^*(z)=\phi_0 - \frac{\Delta \Phi}{2} \tanh\left(z/\xi\right)
\end{align}
where $\xi=\sqrt{\kappa/V_i''}=\sqrt{\kappa/(2r)}$ is the correlation length and $\Delta\Phi=2\sqrt{r/u}$. We can evaluate (\ref{DefGamma})  and obtain
\begin{align}
V_1''=V_2''=\frac{6\gamma}{(\Delta\Phi)^2\xi}.
\end{align}

\paragraph{How small is $\overline{\phi}_i$?\\}

Let us first evaluate the increase of concentration at equilibrium at altitude $z$:
\begin{align}
\frac{\overline{\phi}_1}{\Delta\Phi}=\frac{\rho_{\mathrm{e}} g z}{V''_1\Delta\Phi} = \frac{\xi  \rho_{\mathrm{e}} g z \Delta\Phi}{ 6 \gamma }= \frac{\xi\ z}{6 L_\mathrm{c}^2}. \label{S35}
\end{align}
For $\Delta T=20K$, with $\xi\sim 10$nm and $z=1$mm, $L_\mathrm{c}=20-80\mu$m, the approximation $ \overline{\phi}_1\ll \Delta\Phi$ is justified. Next, the local decrease of micellar concentration due to surface tension at the drop surface is
\begin{align}
\frac{\vert\overline{\phi}_1\vert_{\ve[r]_s}}{\Delta\Phi}\sim \frac{\gamma }{V''_1(\Delta\Phi)^2 R}\sim \frac{\xi}{R}. \label{EqS36}
\end{align}
Since $\xi\ll R$ is always satisfied we conclude that we can safely assume that $\overline{\phi}_1\ll \Delta\Phi$. \\

\paragraph{Discussion of the quasi-static approximation\\}

The quasi-static approximation will be valid if $D_1$ is large compared to $\partial_t R^2$. We can estimate $D_1$ as
\begin{align}
D_1=\lambda V_1''= \frac{6\lambda \gamma}{(\Delta\Phi)^2\xi}=6K_0/\xi,
\end{align}
where $K_0$ is such that $R^3=6K_0(t_\mathrm{f}-t)$ in the surface-tension dominated regime. Let us evaluate the validity of the quasi-stationary approximation in this regime:
\begin{align}
\frac{R\dot{R}}{D_1}\sim \frac{K_0^{2/3} \xi}{(t_\mathrm{f}-t)^{1/3}K_0}\sim \frac{\xi}{R},
\end{align}
meaning that, as long as $\xi\ll R$, the quasi-stationary approximation is well justified. In the gravity-dominated evaporation regime, we have
\begin{align}
\frac{\partial_t R^2}{D_1}=\frac{a}{D_1}\sim \frac{\rho_{\mathrm{e}} g z_{\mathrm{d}}}{\Delta \Phi V''_1}\sim \frac{\xi\ z_{\mathrm{d}}}{6 L_\mathrm{c}^2},
\end{align}
which is small compared to $1$, so that in this regime the quasi-static hypothesis is justified. 

Last, in our experiments it is clear that $\vert\dot{z}_\mathrm{d}\vert<\dot{R}$, so that the above estimates imply that the Peclet number $\text{Pe}=\dot{z}_\mathrm{d}R/D_1$ is also very small compared to unity, justifying the omission of advection terms in the equation for $\overline{\phi}_i$. \\

\paragraph{Effect of quasi-incompressibilty\\}

We now discuss the incompressibility approximation $\nabla\cdot\ve[u]=0$ which is not exact for binary liquid mixtures~\cite{lowengrub1998quasi}, in particular at the interface since the two phases $\Phi_1$ and $\Phi_2$ have different densities. Indeed, mass conservation reads
\begin{align}
\partial_t\rho+\nabla\cdot (\rho\ \ve[u])=0 \label{492},
\end{align}
where $\rho$ is the mass density, $\rho=\phi \rho_{\text{micelles}}+(1-\phi)\rho_{\text{solvent}}$. As already estimated, the variations of $\phi$ inside one phase are negligible compared to $\Delta \Phi$, so that one can consider that the fluid is incompressible inside each phase. At the interface, (\ref{492}) leads to the equation
\begin{align}
&(\Phi_1 \rho_{\text{e}}+\rho_{\text{solvent}})(u_1^\perp-v_\perp)=\nonumber\\
&\phi \rho_{\text{micelles}}+(\Phi_2 \rho_{\text{e}}+\rho_{\text{solvent}})(u_2^\perp-v_\perp),
\end{align}
where $u_i^\perp$ is the perpendicular component of the flow in the phase $i$. 
For example, at the surface of a drop, in the regime where $z_{\mathrm{d}}$ does not vary any longer, the flow induced by evaporation is spherically symmetric, vanishing at the interior ($u_2^\perp=0$), whereas the flow at the exterior of the drop reads
\begin{align}
u_1^\perp=\dot{R}\frac{\Delta\Phi\  \rho_\text{e}}{\Phi_1 \rho_{\text{e}}+\rho_{\text{solvent}}}. 
\end{align}
With typical experimental values at $\Delta T=20$K, we find that $\vert u_1^\perp/\dot{R}\vert < 0.03$. We thus conclude that taking quasi-incompressibility into account in Eq.~(\ref{EqCons}) would  only slightly modify the dynamics of $R(t)$. 

\begin{figure}[h!]
\centering
\includegraphics[width=0.9\linewidth,clip]{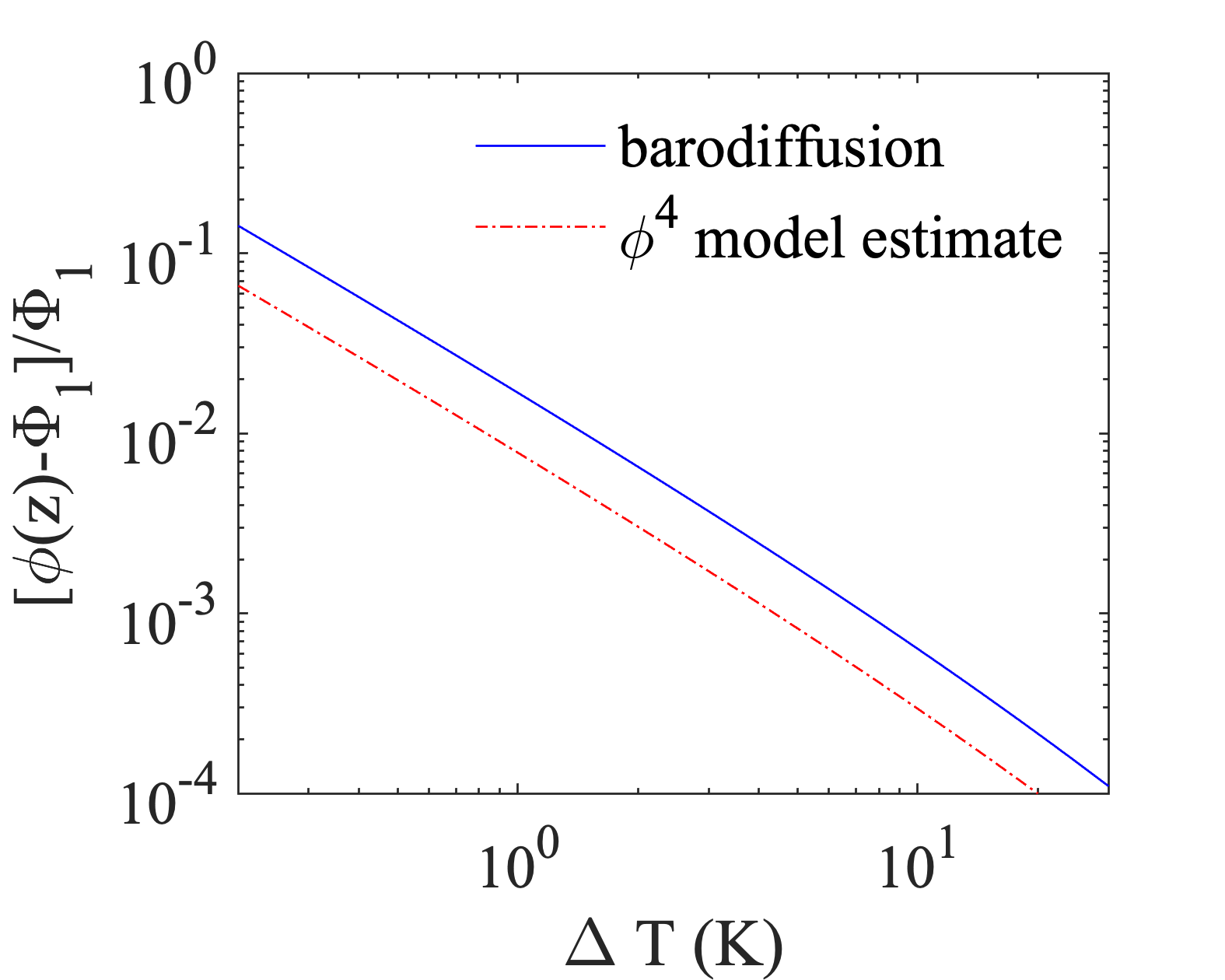}   
\caption{Estimate of the increase of micelle concentration due to gravity in the phase $\Phi_1$, using the barodiffusion theory [blue line, Eq.~(\ref{BarodiffEstimate})] and the $\phi^4$ model [red dash-dotted line, Eq.~(\ref{Phi4Estimate})], for a layer thickness $z=1$ mm.}
\label{FigGradient}
\end{figure}  

\section{ Estimation of the concentration gradient}
The model makes use of the gradient of concentration within the phases induced by   gravity. This effect is related to the so-called barodiffusion (gravity-induced concentration gradient) in binary liquid mixtures [see the chapter ``diffusion'' of  Ref.~\cite{Landau1987}], known to diverge close to a critical point \cite{giglio1975optical,hicks1997thermally}. According to the theory of barodiffusion, a vertical density gradient is necessarily produced in order to keep the system in thermodynamic equilibrium in the presence of gravity. In stationary conditions, this gradient in the separated phase of micellar volume fraction $\Phi_1$ of interest is given in linear conditions by:
\begin{align}
\nabla \phi = \frac{\rho_1}{\rho_{\text{micelles}}}\left(\frac{\partial\rho}{\partial\phi}\right)_T  \chi_T^- \ g \ \hat{\ve[e]}_z,
\end{align}
where $\rho_1$ is the density in the phase $1$, and $ \chi_T^-$ is the osmotic susceptibility in the phase-separated state, as given by Eq.~(\ref{ValueChi}). Since $\rho=\rho_{\text{micelles}}\phi+(1-\phi)\rho_{\text{solvent}}$, we have 
\begin{align}
&\left(\frac{\partial\rho}{\partial\phi}\right)_T=\rho_{\text{micelles}}-\rho_{\text{solvent}}=\rho_\text{e}, \\
&\rho_1\simeq \rho_\text{solvent}+\rho_\text{e}\phi\simeq \rho_\text{solvent},
\end{align}
where the last approximation is justified by $\phi\ll 1$ and $\rho_\text{e}\ll\rho_\text{solvent}$. Collecting these expressions, the variation of $\phi$ in the phase $1$ along a distance $z$, relative to $\Phi_1$, is given in the barodiffusion theory by 
\begin{align}
\frac{\phi(z)-\Phi_1}{  \Phi_1}\simeq  \frac{\rho_\text{solvent} \rho_\text{e} }{\rho_{\text{micelles}}\Phi_1}\   g\  z \ \chi_0^-  \left(\frac{T_C}{T-T_C}\right)^{3\nu-2\beta}.
\label{BarodiffEstimate}
\end{align}
Fig.~\ref{FigGradient}  shows this estimate for $z=1$mm (the thickness of the phase $\Phi_1$), using the numerical values given in Section \ref{SectionNearCritical}. The increase of micelle concentrations becomes significant when $T\to T_c$,  which illustrates why barodiffusion needs to be looked at in near-critical conditions. 
Interestingly, in the $\phi^4$ model one obtains from Eq.~(\ref{S35})
\begin{align}
\frac{\phi(z)-\Phi_1}{\Phi_1} \simeq   \frac{\xi_0^-  \rho_{\mathrm{e}} g z (\Delta\Phi_0)^2 }{ 6 \gamma_0 \Phi_1}  \left(\frac{T_C}{T-T_C}\right)^{3\nu-2\beta}.\label{Phi4Estimate}
\end{align}
This estimate is about two times smaller than that based on barodiffusion theory. So, while diverging close to the critical point, at $\Delta T=20$K, the increase of micelles density is of the order of $0.5\times 10^{-3}\Phi_1$  and seems to be very small. However, gravity plays a significant role when the increase of density is large compared to the increase of concentration due to the Gibbs-Thomson relation, which is estimated in  Eq.~(\ref{EqS36}) to be of the order $\xi\Delta\Phi/R$, which is itself very small compared to one. Hence, the fact that gravity induces only small variations of micelle concentrations with respect to their equilibrium values does not mean that its effect on the droplet decay dynamics is negligible.

\bibliographystyle{naturemag}

\end{document}